\documentclass{article}

\usepackage[final,nonatbib]{neurips_2024}

\usepackage[utf8]{inputenc} 
\usepackage[T1]{fontenc}    
\usepackage{hyperref}       
\usepackage{url}            
\usepackage{booktabs}       
\usepackage{amsfonts}       
\usepackage{nicefrac}       
\usepackage{microtype}      
\usepackage{xcolor}         
\usepackage{graphicx}       
\usepackage{subfigure}

\title{A dynamic parallel method for performance optimization on hybrid CPUs}

\author{%
Luo Yu$^{1*}$ \quad Liu Yucheng$^{1*}$ \quad Shen Haihao$^1$ \quad\\
$^1$Intel Corporation \quad\\
\texttt{\{yu.luo,yucheng.liu,haihao.shen\}@intel.com}\\
}

\begin{document}

\maketitle

\begin{abstract}
  The AIPC concept is gaining popularity, and more and more hybrid CPUs will be running AI models on client devices. However, the current AI inference framework overlooks the imbalanced hardware capability of hybrid CPUs, leading to low inference performance. To address this issue, we have introduced a dynamic parallel method for hybrid CPUs, which significantly increases LLM inference performance by balancing the workload for each core of a hybrid CPU before the parallel work starts. This method has enabled Neural Speed to achieve more than 90\% (on average) of memory bandwidth on two hybrid Intel CPUs.
\end{abstract}

\section{Introduction}

Many CPU companies are now adopting a hybrid CPU architecture to strike a balance between performance and energy consumption. A hybrid CPU typically has more than one microarchitecture or multiple working frequencies among its cores. Both types of hybrid CPUs share a common feature - an imbalance in hardware capabilities among the cores.  For instance, the Intel Ultra 100 Series and Ryzen AI 300 Series have two microarchitectures \cite{intel-aipc}, \cite{amd-aipc}. while the Qualcomm Snapdragon X Elite operates with two frequency levels \cite{qualcomm-aipc}.
These new CPUs are specifically designed to handle large AI models, especially LLM models. 

Recent LLM quantization work shows that 4-bit weight-only quantization is both accurate and efficient for LLM models \cite{inc}, \cite{dettmers2023qloraefficientfinetuningquantized}, \cite{frantar2023gptqaccurateposttrainingquantization}, \cite{cheng2024optimizeweightroundingsigned}. The research also shows that the token generation speed is related to the device's memory bandwidth \cite{wu2023understandingint4quantizationtransformer}. It allows many client CPUs to run some LLM models faster than a human reading speed of about 200ms per token \cite{BRYSBAERT2019104047}.  It's also valuable to continue optimizing the performance of a CPU since it shares the same system bandwidth as its high-performance NPU and GPU.

There are many great works about LLM inference optimization on CPUs. Many of them aim at the server CPUs that are not hybrid \cite{ps2024inferenceaccelerationlargelanguage} \cite{he2024inferenceperformanceoptimizationlarge}. 
$llama.cpp$ \cite{llamacpp}  is a popular low-bit LLM inference framework with good performance on many client CPUs, including hybrid CPUs. Nonetheless, its performance on hybrid CPUs does not meet the performance projection based on the CPU's hardware specification. Although previous research has optimized $llama.cpp$'s x86 CPU codes to achieve higher performance, it still falls short\cite{shen2023efficientllminferencecpus}. Both works utilize the traditional parallel method, resulting in high-performance cores waiting for low-performance cores.

To balance the workload among all cores, it is essential to understand the inference process of the LLM model, which consists of numerous kernels, each with a set of complex instructions. As the throughput of each instruction differs on different cores and the core's performance also varies from platform to platform, it is crucial to dynamically update the core performance and dispatch the kernel problem.

Although OpenMP declares that its parallel\_for algorithm can dispatch the problem from core to core dynamically \cite{omp}. It uses a work-stealing algorithm \cite{work-stealing} and range stealing to balance these partitions when workloads are unbalanced. But, the existing LLM inference framework, like $llama.cpp$, has precomputed and fixed partitions for each thread. It's not easy to change them to parallel\_for style. On the other hand, splitting a matrix multiplication problem into small partitions is not regarded as beneficial. We propose to optimize only the task scheduling part of $llama.cpp$. 

In this paper, we introduce a new dynamic parallel method for LLM inference within the framework of $llama.cpp$. We utilize Neural Speed \cite{ns} which highly optimizes x86 assembly codes based on the work of $llama.cpp$. Our primary contribution is the integration of our new parallel method into Neural Speed, allowing us to achieve more than 90\% of memory bandwidth utilization on two hybrid CPUs during 4-bit LLM inference.

\section{Approach}

In this section, we introduce the two main components of our dynamic parallel method: a CPU runtime and a thread scheduler.

\paragraph{CPU runtime} The CPU runtime records and manages the CPU status. Its thread pool binds each thread to a physical core and it tracks the execution time of each thread during executing kernels. 

\paragraph{Thread scheduler} The scheduler gets CPU status from the CPU runtime and divides kernel tasks into sub-tasks by each core's dynamic performance.

Figure 1 shows our dynamic process during LLM inference: the core performance table is updated after each layer's kernel execution. The engine will adjust the kernel workload among cores through the performance table and maximize the inference performance.
\begin{figure}
    \centering
    \includegraphics[width=0.5\linewidth]{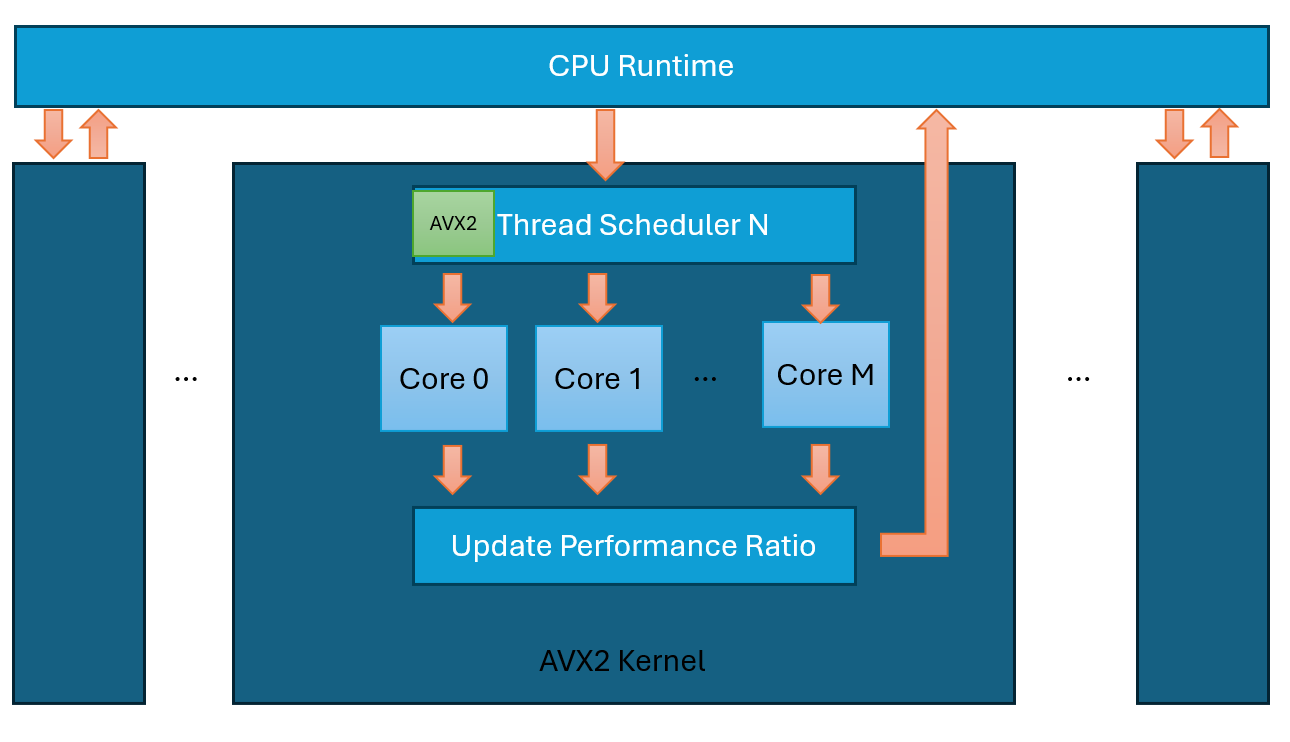}
    \caption{The dynamic LLM inference process}
    \label{fig:process}
\end{figure}

Assume there is a problem with problem size $K$, and there is a hybrid CPU with $N$ cores. We define the performance ratio of i-th core as $pr_{i}$. $pr_{i}$ may be different from each other in a hybrid CPU. The problem will be solved in parallel. Each core will solve a sub set of problem $ps_{i}=\theta_{i}\ K$, and $\theta_{i}\in\left(0,1\right) \ sum\left(\theta_{i}\right)=1$. The execution time of core $i$ is $t_{i}=\frac{ps_{i}}{pr_{i}}$. The time to solve this problem T depends on the slowest core, $T=\max{(}t_{i})$. 
Obviously, the best $\hat{\theta}$ to make the problem solved fastest is  
\begin{equation}
\widehat{\theta_{i}}=\mathop{\arg\min}\limits_{\theta_{i}}{\left(\max{\left(\frac{\theta_{i}\ K}{pr_{i}}\right)}\right)}=\frac{pr_{i}}{\sum{pr_{i}}}
\end{equation}
However, the ${pr}_{i}$ is determined by core frequency, CPU configuration, and even the system background program. We cannot get a static $pr_{i}$ table before runtime.

\subsection{CPU Runtime}

CPU runtime is the runtime context of the CPU. It records the relative performance ratio of each core. We set all ratio values in the performance table at the initialization phase to 1: \(pr_{i}=1\). During runtime, the CPU updates the core performance ratio based on the Instruction Set Architecture (ISA), such as AVX2 and AVX-VNNI for Intel Ultra-125H. Because the P-cores (Performance) and E-cores (Efficiency) of Intel Ultra-125H have different instruction throughput and runtime frequency, it is important to note that different ISAs should have varying performance ratios. Saving performance ratios for each kernel is preferable as a kernel typically utilizes more than one set of instructions. Our experiments have revealed that many kernels share the same performance ratio. As a result, we've designated a primary ISA for each kernel in our paper. 

During LLM inference, the performance ratio will be dynamically updated to adjust the input problem size. If the current performance ratio of the i-th core is \(pr_{i}\) and the calculation time of i-th core is \(t_{i}\) in this calculation. The updated performance ratio ${pr_{i}}'$ is:
\begin{equation}
    {pr_{i}}'=\frac{pr_{i}}{\sum_{j}{t_{i}pr_{j}/t_{j}}}
\end{equation}

Considering the noise data, we use a filter to make our performance table more robust. The performance ratio is finally updated as this equation: $pr_i=\alpha\ast\ pr_i+\left(1-\alpha\right)\ast {pr_i}'$ where $\alpha$ is the constant filter gain.

\subsection{Thread scheduler}

The Scheduler manages parallel problem dispatch for each computing kernel during the inference process, such as tensor copying and matrix multiplication. Its goal is to appropriately dispatch tasks among threads and ensure that all threads complete their sub-tasks simultaneously, even if the performance of each core varies. This approach maximizes CPU performance.

When a kernel involves parallel computations, a scheduler is created, and the ISA primarily used for these computations is specified in the code. The schedule queries the corresponding performance ratios from the CPU runtime based on the current ISA. The scheduler allocates tasks to each thread along a specific dimension according to the performance ratio. Each thread’s sub-task proportion is equal to the ratio of its corresponding core’s performance to the total performance. If the length of dimension is \(s\), and the relative performance ratio of each core is \(pr_{i}\), the sub-task of the i-th core \(s_{i}\) is 
\begin{equation}
    {s_{i}}'=\frac{pr_i}{\sum pr}\ s\
\end{equation}

Afterward, the scheduler activates the thread pool's interface to initiate the threads. Once all sub-tasks of the threads are finished, the scheduler records the execution time for each thread (core) from the thread pool and updates the CPU runtime's performance ratios.

The performance ratio will be distributed among different schedulers. This approach has the advantage of allowing the program to quickly adapt to the current computer conditions and maximize CPU performance, whether during program startup or when there are sudden changes in the system background.

\section{Results}

\subsection{Experiment Setup}
\paragraph{Hardware} We evaluate the performance of two hybrid CPUs: Core-12900K and Ultra-125H. Both are hybrid CPUs with AVX-VNNI instruction.

\paragraph{Baselines} 1. We compare the performance change of INT8 GEMM (General Matrix Multiply) between before and after our integration. We selected the AVX-VNNI GEMM micro kernel of Neural Speed to compute the GEMM problem in our test. We select OpenMP parallel method as a baseline. Please note that OpenMP here uses the balanced work dispatch algorithm. Each thread of OpenMP computes the same size of sub-matrix. We compare the latency of a GEMM problem case and a GEMV (General Matrix-Vector Multiplication) problem case. 2. We compare our model-level performance with Neural Speed and $llama.cpp$. We also select $llama.cpp$ as a baseline because its performance is known by more researchers.

\paragraph{LLM Model} To demonstrate a stable result, we select llama2-7B in 4-bit weight-only quantization as it's widely used. The equivalent data type in $llama.cpp$ is Q4\_0. It uses a group size of 32 for quantization, each group has 32 INT4 data and a FLOAT16 scale. 

\paragraph{Metric} Large Language Model (LLM) inference consists of two distinct phases - prefill phase which processes the input prompt and decode phase which generates output tokens autoregressively\cite{agrawal2023sarathiefficientllminference}. We will use the latency of the prefill phase to measure computing efficiency and the latency of the decode phase to measure memory bandwidth utilization.  

\subsection{Experiment Result}
\begin{figure}[htbp]
    \centering
    \subfigure{
        \includegraphics[width=0.45\linewidth]{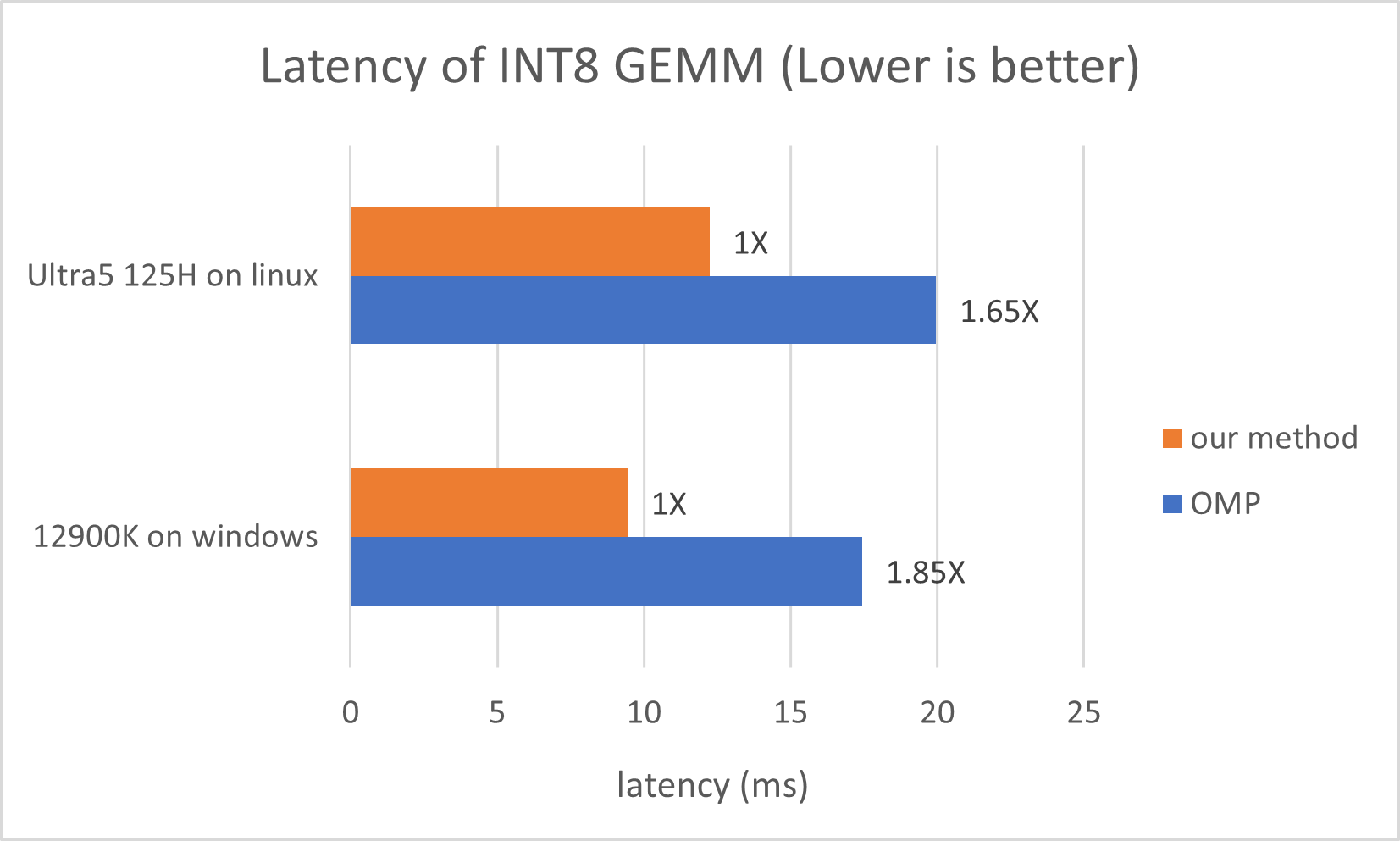}
    }
    \subfigure{
        \includegraphics[width=0.45\linewidth]{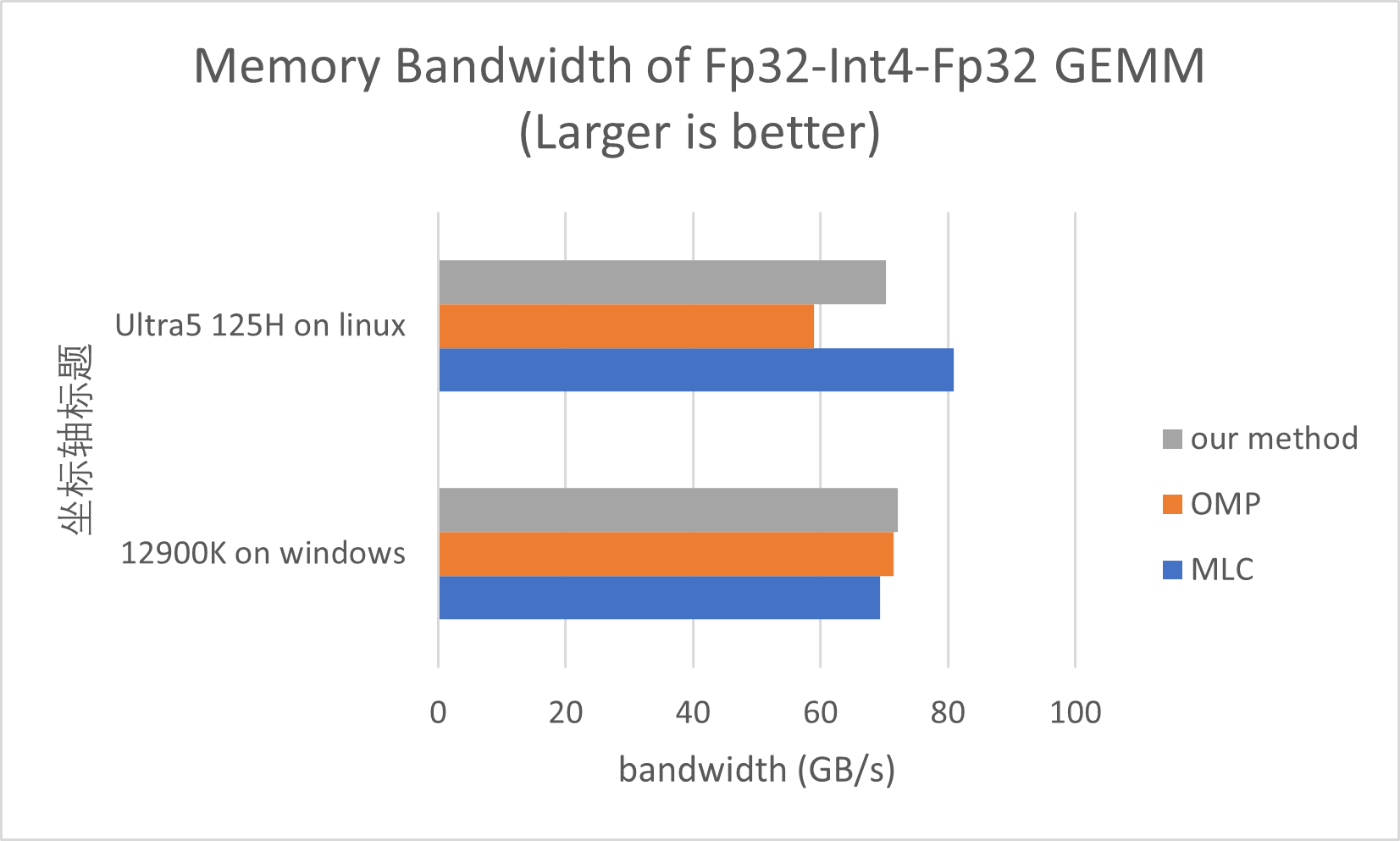}
    }
    \caption{The latency and bandwidth of GEMM in different parallel methods}
    \label{fig:kernel_bench}
\end{figure}

Firstly, we tested the performance of INT8 GEMM. The shape of GEMM is 1024x4096x4096, it usually occurs in the prefill phase. The data type of activation is unsigned INT8. The data type of weight is signed INT8. The data type of output is signed INT32. This GEMM problem is a compute-intensive workload. Our method demonstrates a 65\% improvement in compute performance on Ultra-125H and an 85\% enhancement on Core-12900K. This indicates that our method effectively utilizes the CPU's performance and enhances the collaboration of hybrid cores.

We also tested the performance of INT4 (Fp32-Int4-Fp32) GEMV. Unlike INT8 GEMM, this GEMV includes dynamic quantization for the FLOAT32 input tensor and dequantization for the FLOAT32 output tensor. This represents the complete computation of $llama.cpp$ and Neural Speed. The shape of GEMV is 1x4096x4096. This shape is typical in the decode phase. In this test, we use MLC (Intel® Memory Latency Checker), a tool used to measure memory bandwidth, as a memory bandwidth reference. We convert the kernel's latency data to a memory bandwidth number, as shown in Figure 2. Our method shows a 19\% bandwidth improvement on Ultra-125H. It achieves more than 90\% memory bandwidth of the value MLC measured. 

Secondly, we tested the performance of end-to-end LLM inference. We use two INT4 models of Neural Speed and $llama.cpp$. They are quantized with the same quantization parameters. It means that they have the same model size. The input prompt length is 1024. In this scenario, the prefill phase encountered a computational bottleneck, while the decode phase encountered a memory bandwidth bottleneck. This test shows improved computational efficiency and memory bandwidth utilization with our dynamic parallel method.
\begin{figure}[htbp]
    \centering
    \subfigure{
        \includegraphics[width=0.45\linewidth]{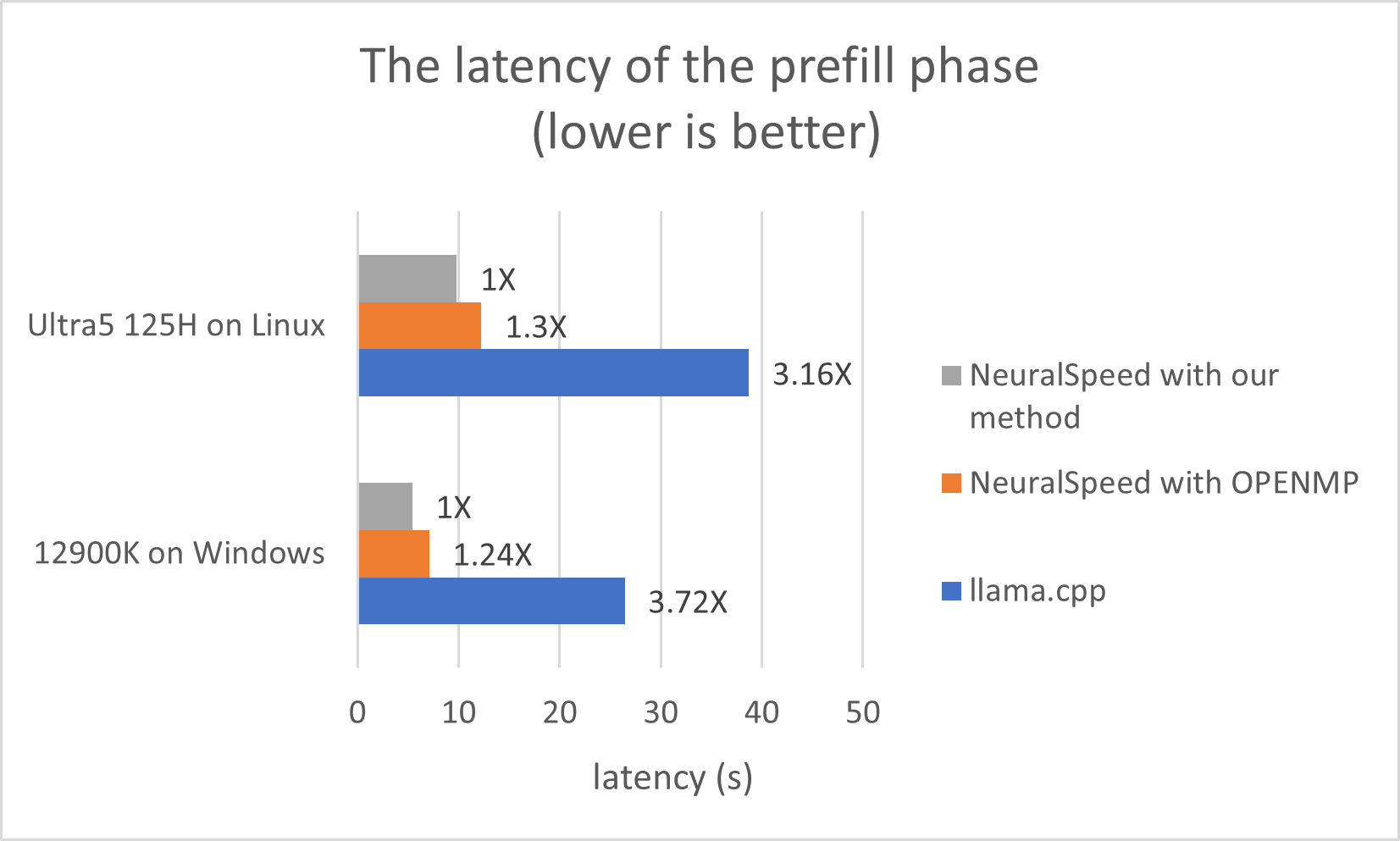}
        \label{fig:sub1}
    }
    \subfigure{
        \includegraphics[width=0.45\linewidth]{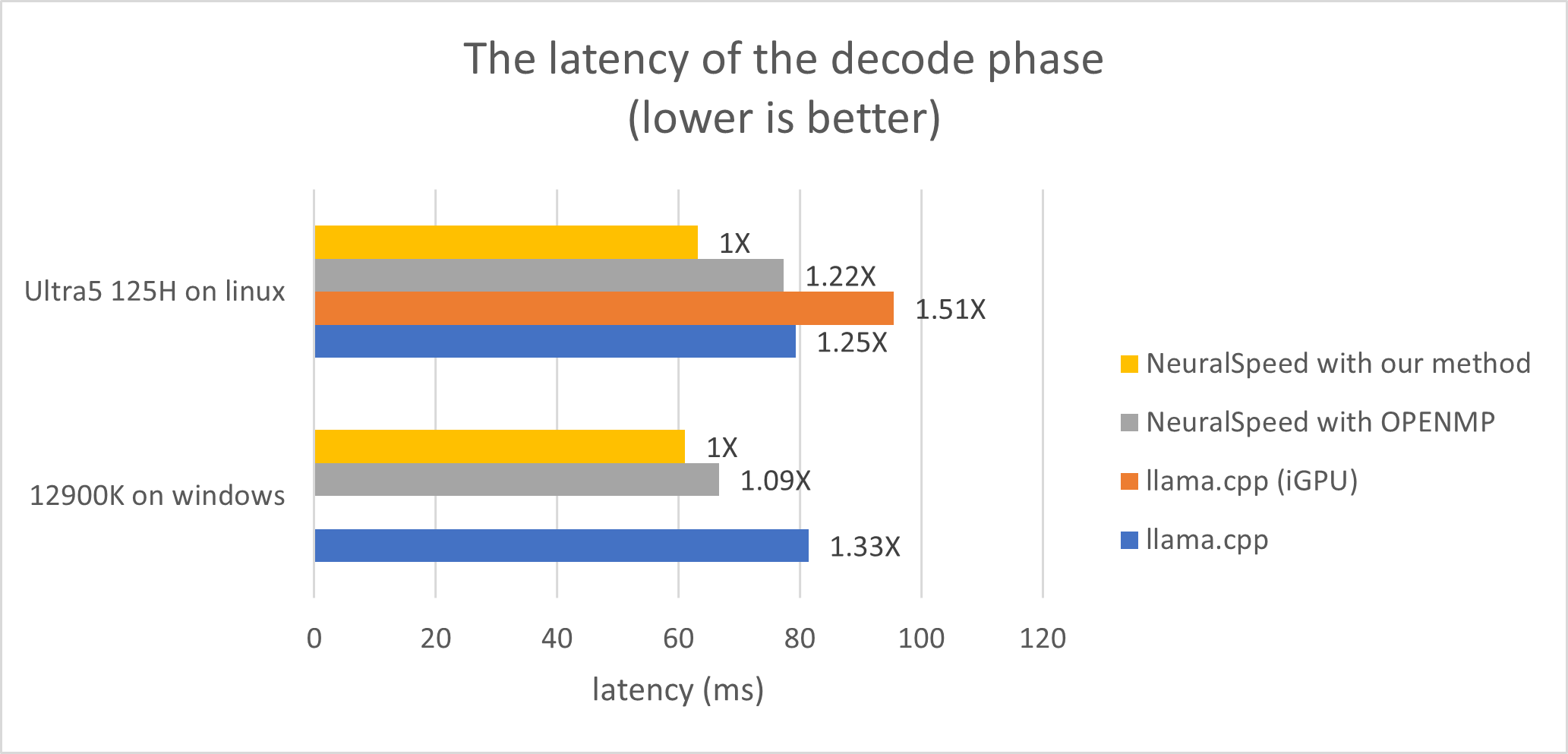}
        \label{fig:sub2}
    }
    \caption{The latency of the prefill phase and the decode phase in Neural Speed (OpenMP and our method) and $llama.cpp$}
    \label{fig:bench_latency}
\end{figure}
In terms of the latency of the prefill phase, our method represents an overall improvement of 20\%-30\% over the original OpenMP method in Neural Speed. One reason that the improvement is lower than the kernel's improvement is that we only apply our method to GEMM kernels. Other kernels, like multi-head attention, do not benefit from our test. Our approach could significantly enhance the compute efficiency of two hybrid CPUs in compute-intensive workloads. As for the latency of the decode phase, our method is 9\%-22\% faster than the original OpenMP method in Neural Speed. The CPU decode speed is about 16 tokens/s.

We also examined how the performance ratios fluctuate during LLM inference. We documented the performance ratios of AVX-VNNI in Ultra-125H's P-cores as displayed in Figure 4. The constant filter gain $\alpha$ is 0.3.

\begin{figure}[htbp]
    \centering
    \includegraphics[width=0.5\linewidth]{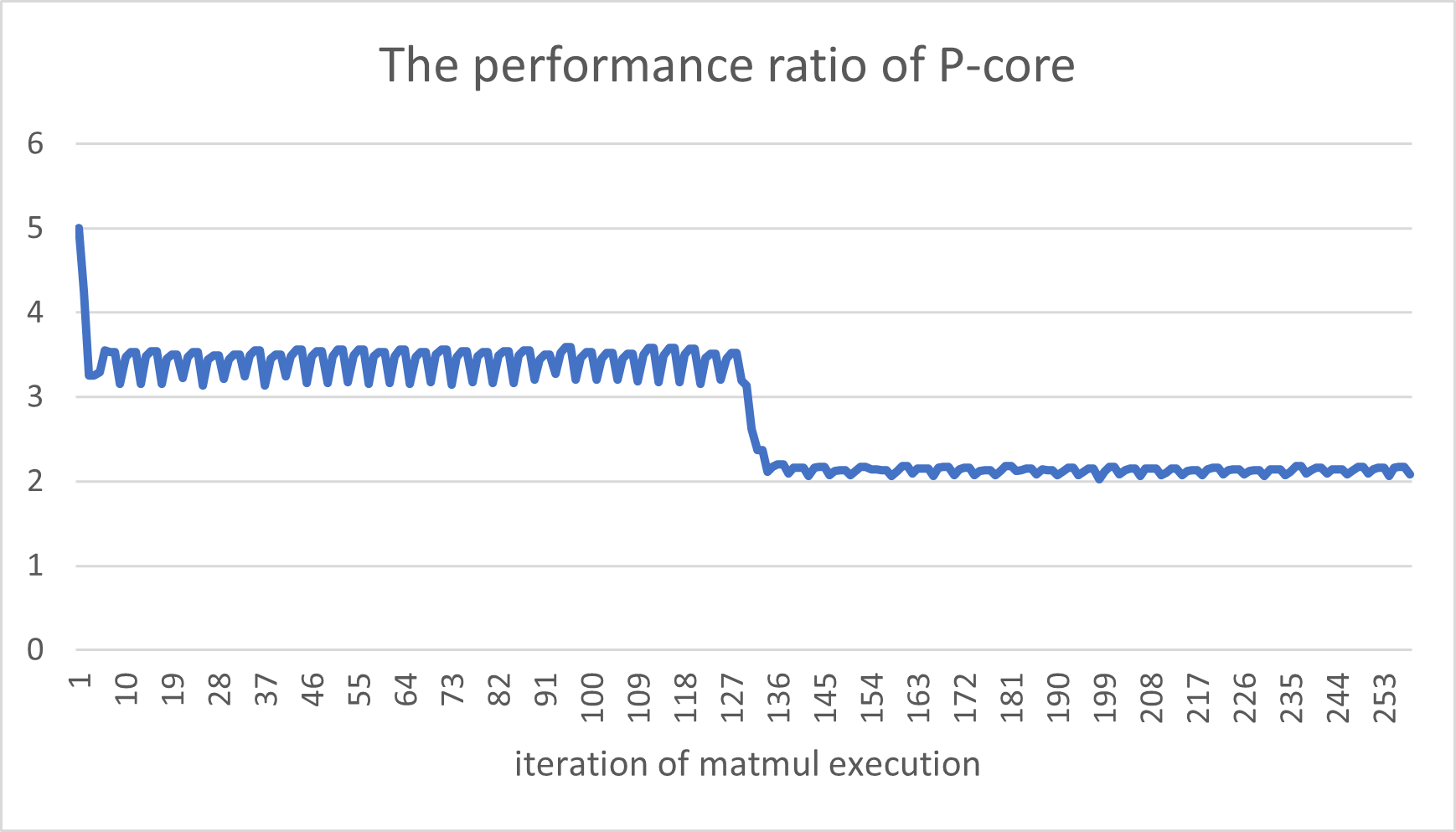}
    \caption{The performance ratio of one P-core in the prefill phase and the decode phase}
    \label{fig:perf_ratio}
\end{figure}
There were two noticeable changes in the performance ratio, as depicted in the picture. The first change occurred after the program started. The performance ratio was initially set at 5, which was too high for this machine, but it quickly stabilized between 3 and 3.5. The second change occurred at the end of the prefill phase and the beginning of the decode phase. The inference of the prefill phase and the decode phase encountered different bottlenecks, resulting in different performance ratios. This test showed that our method can rapidly adapt to varying situations.

The data in Figure 4 explains the difference in speedup between the performance of the prefill phase and the decode phase. The larger the ratio, the greater the differences distributed between different cores in our method. Using the OpenMP method, the P-core would have to wait for the E-core for a longer time if the ratio is higher. Our method can dynamically find these ratios for different hybrid CPUs.

\section{Summary and Future Work}
Our approach enhances LLM inference performance on hybrid CPUs. The new parallel method replaces the OpenMP method of Neural Speed and shows promising performance improvements. With our help, Neural Speed achieves up to a 3.7x speedup compared to $llama.cpp$ and utilizes more than 90\% of the memory bandwidth for INT4 GEMV. Under this memory bandwidth, the two CPUs in our test can generate tokens at about 16 tokens/s. The latest AIPC CPUs have additional compute units, such as GPU and NPU, allowing us to dynamically dispatch an LLM kernel between these units to utilize them simultaneously. Our next objective is to optimize the hybrid compute units to decrease the latency of the prefill phase on AIPC.
\bibliographystyle{plain}
\bibliography{neurips_2024}

\end{document}